\documentstyle[aps,epsf,twocolumn]{revtex}

\def\la{\langle}
\def\ra{\rangle}

\def\Lsu2{{\cal L}_{\mbox{SU(2)}}}

\def\su2xsu2{{SU(2)\times SU(2)}}
\def\su3xsu3{{SU(3)\times SU(3)}}

\def\be{\begin{eqnarray}}
\def\ee{\end{eqnarray}}
\def\ben{\begin{eqnarray}}
\def\een{\end{eqnarray}}

\newcommand{\beq}{\begin{equation}}
\newcommand{\eeq}{\end{equation}}
\newcommand{\bea}{\begin{eqnarray}}
\newcommand{\eea}{\end{eqnarray}}

\begin{document}

\draft
\title{\bf  Chiral Random Matrix Models :\\
\bf Thermodynamics, Phase Transitions and Universality }

\author{ {\bf Romuald Janik}$^1$, {\bf Maciej A.  Nowak}$^{1,2}$ 
and {\bf Ismail Zahed}$^3$}

\address{$^1$ Department of Physics, Jagellonian University, 
30-059 Krakow, Poland.\\
$^2$ GSI, Plankstr. 1, D-64291 Darmstadt, Germany \& \\
 Institut f\"{u}r Kernphysik, TH Darmstadt, D-64289 Darmstadt, Germany \\
$^3$Department of Physics, SUNY, Stony Brook, New York 11794, USA.
}
\date{\today}
\maketitle

\begin{abstract}
For one flavour, we observe that standard chiral random matrix models 
are schematic variants of the 
Nambu-Jona-Lasinio (NJL) models whether in vacuum or matter.  The ensuing 
thermodynamics is that of constituent quarks, with mean-field universality
in general.
For two and three flavours non-standard chiral random matrix models with 
$U_A(1)$ breaking are suggested. For three flavours the transition is
reminiscent of the isotropic-nematic transition in liquid crystals.

\end{abstract}

{\bf 1.\,\,\,}
A large number of finite-temperature and density studies in the recent past 
have focused on the Nambu-Jona-Lasinio model (NJL) or extension thereof
\cite{NJL}. These models capture the essentials of the spontaneous breaking of 
chiral symmetry in the vacuum, and have been used to model the properties of
the low-lying pseudoscalar nonet. They bear much in common with more 
microscopically motivated thermodynamical descriptions of the QCD vacuum such 
as the instanton model \cite{INS}.

In these models, the low-lying 
spectrum is composed of constituent quarks and mesons. In the ground state, 
the quark-antiquark interaction is attractive in the singlet-isosinglet 
providing a simple mechanism for the spontaneous breaking of chiral symmetry,
through quark condensation. With increasing density or temperature, the 
constituent quarks in matter overcome the "asymmetry" produced by 
the condensation of the quarks, leading to a chirally symmetric phase of 
screened quarks. The transition is mean-field in nature, and mostly entropy 
driven. 

Since chiral symmetry breaking is 
encoded in the eigenvalue distribution at zero virtuality, it is clear that a 
chiral phase transition would affect quantitatively this distribution. 
Chiral random matrix models have proven to be a very efficient way to
get at the eigenvalue distribution of  QCD inspired Dirac spectra. In this 
letter, we investigate the relationship between chiral random matrix models and 
Nambu-Jona-Lasinio models both in vacuum and matter. In section 2, we make this 
relation quantitative. In section 3 we discuss the thermodynamics in the 
mean-field approximation. In section 4, we  address some issues related to 
the Dirac spectrum in the mean-field approximation. 
In section 5, we discuss new and non-standard random 
matrix models with $U_A(1)$ breaking as inspired by effective models. In 
section 6, we comment on how the issue of confinement may affect the 
thermodynamical arguments. Our conclusions are summarized in section 7.

\vskip .5cm
{\bf 2.\,\,\,}
To illustrate these points, consider first the case of one flavour, with the 
schematic Lagrangian in four Euclidean space as given by
\be
{\cal L}_4 = &&+{\psi}^{\dagger}( i\gamma\cdot \partial  + im + i\mu \gamma_4 )
\psi \nonumber\\ &&+ \frac {g^2}{2}
\bigg( (\psi^{\dagger}\psi )^2 + (\psi^{\dagger} i\gamma_5 \psi)^2 \bigg)
\label{PT1}
\ee
or equivalently 
\be
{\cal L}_4 =&&+ {\psi}_R^{\dagger}( i\gamma\cdot \partial   + i\mu \gamma_4 )
\psi_L +
 {\psi}_L^{\dagger}( i\gamma\cdot \partial  + i\mu \gamma_4 )
\psi_R \nonumber\\
&& + \psi_R^{\dagger}\, i ( P + m) \, \psi_R + \psi^{\dagger}_L i ( P^{\dagger} 
+ m) 
\psi_L  + \frac {1}{2g^2} PP^{\dagger}
\label{PT2}
\ee
in the chiral basis $\psi = (\psi_R , \psi_L )$. Here $P, P^{\dagger}$ stand
 for  independent 
auxiliary fields, $g$ is a fixed coupling and $\mu$ is 
 a real chemical potential.
Note that the Minkowski fields follow from the Euclidean fields through 
$(i\psi^{\dagger} , \psi )\rightarrow (\overline{\psi} , \psi )$. 
Equation (\ref{PT2}) is defined on 
the strip $\beta \times V_3$ in Euclidean space, 
with 
$P(\tau + \beta , \vec x ) = P (\tau, \vec x )$, and $\psi (\tau + \beta , 
\vec x ) = -\psi (\tau, \vec x )$.

To establish the connection to  chiral random matrix models \cite{RANDOM}, 
further simplifications are needed.
The anti-periodicity of the quark fields yields
\be
\psi (\tau , \vec{x} ) = \sum_{n=-\infty}^{+\infty} \,\,e^{-i\omega_n \tau} 
\,\psi_n^x 
\label{matsu}
\ee
where $\omega_n= (2n+1) \pi T$ are the Matsubara 
frequencies ($T=1/\beta $), 
and $x=1,2, ...,N$ label discrete points in space. 
Space is here a grid of dimension 
$N$, where each point contains a  quark of frequency~$n$.
If we further assume that the 
auxiliary fields  $P, \,P^{\dagger}$ are
 constant in space and time, the action in (\ref{PT2})
reduces dimensionally  to a 0-dimensional one 
 with infinitely many Matsubara modes.
The corresponding partition function  can be readily found in the 
form
\be
Z[T,\mu ] = &&\int \, dP\, e^{-N\beta\Sigma  
PP^{\dagger}}\nonumber\\&&\times
\prod_{n=-\infty}^{+\infty} \, {\rm det}_2^N 
 \beta \left( \matrix{i (m + P)& \omega_n +i\mu \cr \omega_n + i\mu & i (m+ 
P^{\dagger}) \cr}\right)
\label{X1}
\ee
following the rescaling 
$q_n^x = \sqrt{V_3} \psi_n^x$ and $\Sigma = {V_3}/{2g^2}$,
where $q_n^x$ are now dimensionless Grassmann variables.
The determinant in (\ref{X1}) is over $2\times 2$ matrices. The corresponding
0-dimensional Lagrangian is
\be
{\cal L}_0 =
+q^{\dagger} ( ({\bf \Omega} + i\mu ) \gamma_4 + im ) q 
 +\frac 1{N\Sigma}  q^{\dagger}_L q_L q^{\dagger}_Rq_R
\label{X44}
\ee
with ${\bf \Omega} = \omega_n{\bf 1}_n \otimes {\bf 1}_x$.

Consider now the new auxiliary matrix ${\bf A}_{n,m}^{x,y}$ with entries both 
in ordinary space $x,y$ and frequency space $n,m$. 
${\bf A}$ is a doubly banded, complex matrix with dimensions
$(N\times N)\otimes (\infty \times\infty )$. In contrast with $P$, 
the matrix ${\bf A}$ bosonizes pairs of quarks of 
{\bf opposite} chirality. For the lowest two Matsubara 
frequencies it is simply $(N\times N)\otimes (2\times 2)$ matrix .
In terms of ${\bf A}$, the analogue of (\ref{X44}) is
\be
{\cal L}_0 =&& 
+q^{\dagger} ( ({\bf \Omega} + i\mu ) \gamma_4 + im ) q \nonumber\\
&& +N \Sigma {\rm Tr}_{x,n} ({\bf A}{\bf A}^{\dagger})
+ q_{R}^{\dagger } {\bf A} q_{L} + q_{L}^{\dagger} {\bf A}^{\dagger}q_{R}
\label{X4}
\ee
The trace in (\ref{X4}) is over $x$ and $n$. The partition function 
associated to (\ref{X4}) is simply
\be
Z[T, \mu ] = &&\int \,\,d{\bf A}\,\, 
e^{-N\beta\Sigma {\rm Tr}_{x,n} ({\bf A}{\bf A}^{\dagger} )}\,\,
{\rm det}_{2,x,n}\,\beta\,{\bf Q}
\label{X5}
\ee
with the medium Dirac operator in a random background,
\be
{\bf Q} = \left(\matrix{ im &  {\bf \Omega} + i\mu \cr
 {\bf \Omega} +i \mu & im \cr}\right) + \left(\matrix{ 0 & {\bf A}\cr
{\bf A}^{\dagger} & 0 \cr}\right)
\label{XDIRAC}
\ee
The determinant in (\ref{X5}) is over chirality (2), space ($x$), and frequency 
space ($n$).
This is an example of a chiral random matrix model 
\cite{RANDOM,JACKSON,STEPHANOV}.

\vskip .5cm
{\bf 3.\,\,\,}
To discuss the thermodynamics of the random matrix model, it is best to use
(\ref{X1}) in large $N$  with ${\bf n}=N/V_3$ fixed. 
The Gibbs free energy associated to (\ref{X1}) reads
\be
\Omega = &&-N \left( \omega +
 T \,{\rm ln} \prod_{\pm} ( 1+e^{-(\omega\mp \mu )/T} )\right)\nonumber\\
&&+N\Sigma PP^{\dagger}
\label{X7}
\ee
with $\omega^2={(P+m)(P^{\dagger}+m)}$. The first term in (\ref{X7})
is just the Hartree contribution to the Gibbs
free energy (Fig. 1a), while the second
term corrects for the double counting of the interaction energy (Fig. 1b),
that is
\be
N\Sigma PP^{\dagger} = -\frac 1{N\Sigma} <q_L^{\dagger} q_L> <q_R^{\dagger} 
q_R>
\ee
In this schematic model, the Dirac spectrum is simplified to two-levels in frequency 
space ($\pm \omega$) for each $x=1,2, ...,N$. The quark fields do not carry 
spin. Recall that $N_F=1$. There is no kinetic energy 
associated to the quarks in either (\ref{X1}) or (\ref{X5}). These degrees
of freedom are also manifest in the pressure at high temperature,
\be
{\bf P} = -\frac{\Omega}{V_3} =
2{\bf n}T {\rm ln} 2  
- {\bf n}\Sigma PP^{\dagger} +{\cal O}(\frac{1}{T})
\label{X8}
\ee
where $2{\bf n}$ is the number of quarks and antiquarks, and ${\rm ln2}$ their
respective entropy at $T=\infty$ since their occupation number is $1/2$.
The $1/T$ term in the entropy cancels against the vacuum energy in the 
Hartree contribution. The first term in (\ref{X8}) is the thermal pressure 
of  free constituent  quarks, while the second term is the left out 
interaction energy.

\begin{figure}
\centerline{\epsfxsize=7cm \epsfbox{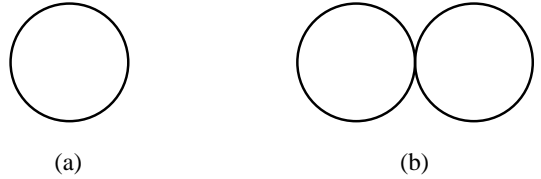}}
\caption{\label{fig1} (a) is the Hartree contribution to the Gibbs free
energy, and (b) removes the overcounting.}
\end{figure}

In large $N$, the extremum of the Gibbs 
free energy (\ref{X7}) yields a gap equation 
for $P_*$ 
\be
2 P_*\Sigma =  1-n-\overline{n} 
\label{X9}
\ee
where $n = (e^{(\omega_* -\mu)/T} +1)^{-1}$ for particles and
$\overline{n} = (e^{(\omega_* +\mu)/T} +1)^{-1}$ for antiparticles.
The constituent quark density is 
\be
\rho_Q= 
i\la \psi^{\dagger} \gamma_4 \psi \ra =
 -\frac{1}{V_3}\frac{\partial \Omega}{\partial \mu}=
   {\bf n}\, (n-\overline{n} )
\label{X10}
\ee
while the quark condensate is
\be
i\la \psi^{\dagger}\psi \ra =
\frac{1}{V_3}\frac{\partial \Omega}{\partial m} =  
- 2{\bf n} \, \Sigma  P_*
\label{X11}
\ee
The relations (\ref{X9}-\ref{X11}) are just schematic versions of the usual 
gap, density and condensate relations in the NJL models. 

For zero chemical potential, (\ref{X9}) admits solutions for $4\Sigma T_* \leq 
1$ in the massless case. For zero temperature, $n=1$ for $\mu \geq\omega$ and 
zero otherwise, and $\overline{n}=0$. For $\mu <\omega$ the pressure of the
constituent quarks is constant ${\bf P}={\bf n}/4\Sigma$, while that of the
free quarks increases linearly with $\mu$, ${\bf P}={\bf n}\mu$. 
This behaviour is not compatible with confinement at small $\mu$ (see section 
6). A first order 
transition to the free phase takes place at $\mu_*=\omega/2$, in which chiral
symmetry is restored. Clearly the quark number density is zero in the 
constituent quark phase (Fermi level in the Dirac gap) and ${\bf n}$ in the 
free phase. Since $\omega$ plays the role of the constituent quark mass,
chiral symmetry is qualitatively restored for $\mu_*\sim m_N/6$, where $m_N$
is the nucleon mass. The extra factor of $1/2$ follows from the peculiar form 
of the pressure in the absence of kinetic energy.

If we were to retain only {\bf few} Matsubara modes in the random matrix model
then the thermodynamical relations are altered {\bf qualitatively}
at zero chemical potential and {\bf quantitatively} otherwise. Indeed, at zero
chemical potential the Helmholtz free energy with one Matsubara mode becomes
\be
F= - \frac N{\beta}{\rm ln} \beta^2 (\omega^2 + \pi^2 T^2)+ N\Sigma PP^{\dagger}
\label{M1}
\ee
in comparison with (\ref{X7}). Setting $\beta=\Sigma$
\footnote{The other choice $\beta=2\Sigma$ amounts for a global shift in the 
pressure, and would not affect quantitatively the discussion.}
to ensure a finite $T\rightarrow 0$ limit, 
yields  the gap equation $\omega_*^2+\pi^2 T^2=\omega_*/P_*\Sigma^2$, for 
which the pressure is ${\bf P}= -{\bf n} \Sigma P_*^2$.
For $\beta=\Sigma$ a single Matsubara mode contributes zero to the Hartree
part of the free energy (Fig. 1a). This is reminiscent of the high temperature 
cancelation mentioned above. We note that due to the choice 
$\beta=\Sigma$ the pressure grows quadratically in this phase as opposed to
linearly in (\ref{X8}). Since the pressure of a free Matsubara mode is
${\bf P}_F= 2{\bf n}{\rm ln} (\Sigma\pi T)/\Sigma$, a phase change at
$\pi\Sigma T_*=1$ is expected, for which ${\bf P}={\bf P}_F=0$. This 
transition is characterized by $i<\psi^{\dagger}\psi > =-2{\bf n} \Sigma P_*$ 
as an order parameter, and was first discussed in \cite{JACKSON}.

In the case where {\bf no} Matsubara mode is retained {\bf and} $\mu$
is finite, the results follow through the substitution $\pi T\rightarrow i\mu$
above. In particular, the constituent quark mass 
$P_*=\sqrt{1+\Sigma^2\mu^2}/\Sigma$ (massless case)
grows with $\mu$. Hence $i<\psi^{\dagger}\psi > 
= -2{\bf n}\Sigma P_*$ increases in
strength, while $\rho_Q = -2{\bf n} \Sigma \mu$ decreases. These
behaviours are counter-intuitive. Indeed, in this phase
the pressure ${\bf P} = -{\bf n}  \Sigma P_*^2$ decreases with increasing $\mu$.
Such a system will tend to cavitate (collapse)
unless a phase transition sets in. 
This is unlikely since the pressure of the free phase is {\bf complex}
\be
{\bf P}_F = 2{\bf n} \frac 1{\Sigma} {\rm ln}\mu\Sigma \pm {\bf n} \frac 
{i\pi}{\Sigma}
\label{M6}
\ee
The free system is unstable. The present analysis suggests that 
the correct thermodynamical behaviour is only maintained
if {\bf all} Matsubara modes are retained in the random matrix model, with the
zero temperature limit taken after the large (three) volume limit.

At zero chemical potential, 
the random matrix models we constructed depend on the number and 
mixing of the various Matsubara modes. At high temperature we expect only 
the lowest modes 
$\omega_0$ and $\omega_{-1}$ to survive, making the various matrix models 
equivalent to each other at the {\bf critical point}.
Indeed, at high temperature and for  $T \approx T_c$ the gap equation 
(\ref{X9}) simplifies to 
\be
g (P_* + m)^3  + \mu^2 (P_*+m) - h = {\cal O} (\frac 1T )
\label{X12}
\ee
with $h=m$, $\mu^2 = (T-T_c)/T_c$ and $g=(12T^2_c)^{-1}$. Eq. (\ref{X12})
has the generic form of a cubic equation, as  expected from mean-field 
treatments
of chiral phase transitions \cite{PISARSKI}. Indeed, (\ref{X12}) is just the 
gap equation generated from the effective potential \cite{PISARSKI}
\be
{\cal L}_0 (T, \Phi )= -h \Phi + \frac 12 \mu^2 \Phi^2 -\frac {c}{N_F} 
\Phi^{N_F} + \frac{g}{4} \Phi^4
\label{Y1}
\ee
where in the present case $c=0$. This observation implies that the chiral 
random matrix model (\ref{X4}) much like the NJL model (\ref{X44})
enjoys mean-field critical exponents
$(\alpha, \beta, \gamma, \delta, \nu, \eta ) = (0, \frac 12, 1, 3, \frac 12 , 
0)$.

\vskip .5cm
{\bf 4.\,\,\,}
To discuss the spectral distribution of the Dirac operator in the medium, it is 
best to use (\ref{X5}-\ref{XDIRAC}), instead of (\ref{X1}). For {\bf one} 
Matsubara mode, the resolvent of (\ref{XDIRAC}) is
\be
{\bf G} (z) = \frac 1N \left< {\rm Tr}_N \frac 1{z-{\bf Q}} \right>
\label{RESOLVENT}
\ee
where the averaging is carried using (\ref{X5}) in the unquenched 
approximation. The spectral distribution associated to (\ref{RESOLVENT})
follows from its discontinuity through the real
axis, $\pi\nu(\lambda ) = -{\rm Im} {\bf G}(\lambda + i0 )$.
In the large $N$ limit and in the quenched approximation, the resolvent
(\ref{RESOLVENT}) can be readily found to satisfy in general
Pastur's equation \cite{PASTUR,ZEE}.
For $m=\mu=0$ and at high temperature (only one Matsubara mode 
retained), Pastur's equation for ${\bf G} (z)$ is a cubic equation of the form
\be
G^3 -2z G^2 + (z^2-\pi^2 T^2  + 1) G -z =0
\label{XCARDANO}
\ee
This equation was first discussed in \cite{BHZ}
in a general context, and used by others \cite{JACKSON,STEPHANOV,BLUE,PAPP}.
The cubic equation (\ref{XCARDANO}) is a direct manifestation of 
(\ref{Y1}) and hence mean-field universality \cite{PISARSKI}. The spectral 
distribution at the critical point following from (\ref{XCARDANO}) is 
consistent with a numerical analysis \cite{JACKSON} using large matrices.

For $T=m=0$ and finite $\mu$ the situation is more subtle. Naively, the 
resolvent in this case satisfies (\ref{XCARDANO}) with
the substitution $\pi T\rightarrow i\mu$. The discontinuities of ${\bf G} (z)$ 
are valued in the z-plane, with end-points given by
the zeroes of the discriminant
\be
4\mu^2 z^4 + z^2 (8\mu^4 - 20 \mu^2 -1) + 4 (\mu + 1)^3 =0
\label{DISCRIMINANT}
\ee
For $\mu=0$ there are two real roots $z=\pm 2$. The result 
is a cut along the real axis between $-2$ and +2. The discontinuity 
along the cut is Wigner's sumrise distribution for the spectral density.  
With increasing $\mu$, the 
two roots approach each other and tend to $\pm \sqrt{27/8}$, followed by the 
emergence of two new roots from real infinity. These new roots
are branch points of cuts that are rejected to infinity (unbound 
spectrum). For $\mu_*^2=1/8$, the branch points coalesce pair-wise on the 
real axis, and for $\mu^2 >1/8$ they move off the real axis resulting
into four endpoints that are symmetric about the real axis. The present 
large $N$ analysis suggests that the spectral density jumps to zero at 
$\mu_*=1/\sqrt{8}\Sigma$
\footnote{The value of $\mu_*\approx\sqrt{0.278}$ quoted in 
\cite{STEPHANOV} uses the real part of a complex pressure.}. 
Since the pressure (\ref{M6}) of the free system 
is complex, this jump is simply {\bf spurious}. Indeed, direct numerical 
analyses \cite{STEPHANOV,STONY} have revealed that the  spectral distribution 
at finite $\mu$ is complex valued whatever $\mu >0$. Technically, it is 
sufficient to note that the complex 
character of the Dirac operator at finite $\mu$ causes large fluctuations in 
the eigenvalue distributions, thereby upsetting the mean-field
approximation for the resolvent in certain parts of the z-plane
\cite{STONY,USNEW}. 

These effects are symptomatic of the fact that the bulk 
pressure shows signs of cavitation (decreasing ${\bf P}$) and unstability 
(complex ${\bf P}_F$). A simple fix would be to include {\bf all} Matsubara 
modes as we indicated above. Since chiral symmetry is only restored for 
$\mu_*\sim m_N/6$ in this case, and the thermodynamical system is stable, 
it would be interesting to investigate the properties of the resolvent
and the Dirac spectrum. 

Quenched lattice simulations \cite{MULATTICE} suggest that the transition
sets in at $\mu >0$. This may mean at least two thinghs : either that the 
lattice simulations do not generate the proper thermodynamical ensemble, or 
that quenched simulations do not follow from the conventional spectrum. The 
former can be checked by simulating the free massive quark ensemble at finite
temperature and chemical potential, and probing the various limits. The latter
has been suggested recently in \cite{STEPHANOV}, where it was argued
that quenched lattice simulations follow from a new spectrum composed of 
quarks $q$ and their conjugates $Q$ (copies of opposite baryon number). 
Specifically, the analog of (\ref{X44}) is now
\be
{\cal L}_0 = &&+q^{\dagger} \gamma_4 i\mu q -Q^{\dagger} \gamma_4 i\mu Q
\nonumber\\ &&+\frac 1{N\Sigma} (q^{\dagger}_Lq_Lq^{\dagger}_R q_R +
Q^{\dagger}_L Q_L Q^{\dagger}_R Q_R ) \nonumber\\
&&+\frac 1{N\Sigma} (q^{\dagger}_LQ_LQ^{\dagger}_R q_R +
Q^{\dagger}_L q_L q^{\dagger}_R Q_R ) 
\label{S1}
\ee
As a result, the pressure of the free phase is now
real  ${\bf P}_F= 4{\bf n} {\rm ln} (\Sigma \mu)/\Sigma$ and monotonously
increasing with $\mu$. This pressure is physically meaningful only for 
$\Sigma\mu\geq 1$, since a free system with negative pressure will cavitate.
Besides the chirally broken phase with $i<q^{\dagger} q>\neq 0$
and an unphysical pressure  (monotonously decreasing and negative),
(\ref{S1}) allows for a 
chirally symmetric phase with a mixed condensate $<Q^{\dagger} q>$ and a 
physical pressure. In the large $N$ (Hartree) approximation, the mixed
condensate satisfies the gap equation
\be
ip_*=&&  +\frac 1{N\Sigma} <Q_R^{\dagger}q_R>\nonumber\\
=&&-\frac 1{4\Sigma^2} {\rm Tr} \left(\tau_1\frac 1{\gamma_4i\mu\tau_3 
+ip_*\tau_1}\right)
\label{S2}
\ee
where $\tau$'s are Pauli matrices active on the doublet $\xi = (q, Q)$. 
Hence $p_*^2+\mu^2=1/\Sigma^2$. The appearance of the mixed condensate, 
is tantamount to the spontaneous breaking of a vector symmetry, that is
$<\xi^{\dagger} \tau_1 \xi>\neq 0$
\footnote{In gauge theories this is not in conflict with the Vafa-Witten 
theorem \cite{VW} since $\mu$ is nonzero.}. 
The effect of the conjugate quarks is to change $\mu^2$ to $-\mu^2$ in the gap 
equation (thanks to $\tau_1$) thereby fixing the pressure at low $\mu$,
\be
{\bf P} = +\frac 2{N\Sigma V_3} <Q_R^{\dagger} q_R>^2 = -2{\bf n}\Sigma p_*^2
\label{S3}
\ee
as expected from Fig. 1b, since Fig. 1a contributes to zero due to (\ref{S2})
for $\beta=\Sigma$. The quark number density in this case is 
$i<\xi^{\dagger}\gamma_4\tau_3\xi>/V_3 = 4{\bf n} \Sigma\mu$. This
behaviour is not compatible with confinement at small $\mu$ (see section 6).
For $\Sigma\mu_* =1$, the quark number density and the pressure are those
of the free constituents, so a phase admixture is expected. Using mean-field 
arguments, the resolvent associated to (\ref{S1}) was discussed in 
\cite{STEPHANOV} in the quenched approximation, and overall consistency with 
the numerically generated Dirac spectrum was found.

\vskip .5cm
{\bf 5.\,\,\,}
For two and more flavours, we can use the above analogy with the NJL model 
to construct non-standard chiral random matrix models with explicit $U_A(1)$ 
breaking. Indeed, (\ref{X1}) suggests the minimal generalization to two 
flavours
\be
&&Z[T,\mu ] = \int \, dP \, dP^{\dagger}\, 
e^{-N\beta\Sigma  ({\rm det}_f P + {\rm det}_f P^{\dagger} )}
\nonumber\\
&&\times\prod_{n=-\infty}^{+\infty} \,\,
{\rm det}_{2,f}^N\beta \left( \matrix{i (m+P) & \omega_n + i\mu \cr
\omega_n + i\mu & i (m + P^{\dagger} ) \cr}\right)
\label{YX21}
\ee
Here $P$ and $P^{\dagger}$ are complex $2\times 2$ valued auxiliary fields,
that transform respectively  as $V_{L} P V_{R}^{\dagger}$ and
$V_{R} P^{\dagger} V_{L}^{\dagger}$ under $U(2)_{L}\times U(2)_R$. 
In the massless case, (\ref{YX21}) is $SU(2)_L\times SU(2)_R$ symmetric.
We note that the $P$-integration can be traded to an integration over
Grassmann variables, leading to a standard two-flavour NJL model with a
determinantal interaction. The QCD motivation for this model can be found
in \cite{NJL,INS} and references therein. A rerun of the above arguments 
allows a rewriting of (\ref{YX21}) into
\be
Z[T,\mu ] = &&\int \, dA_R \, dA_L \, e^{-\beta\Sigma \,\,{\rm Tr}_{xf} (A_R^a A_R^a + 
A_L^a A_L^a )}\nonumber\\
&&\times {\rm det}_{f,x,n}\beta \left(\matrix{im + \tau^aA_R^a&\Omega + i\mu\cr
\Omega + i\mu & im + \tau^a A_L^a\cr}\right)
\label{Y20}
\ee
where $\tau^a= ({\bf 1}, i\vec{\tau} )$ and $a=0,1,2,3$. For each frequency 
$n$, the entries in (\ref{Y20}) are $(2,2)\otimes (N,N)$ valued, with $m$, 
$\Omega$ deterministic and $A_{R,L}$ random and hermitean.

In the large $N$ limit, the parity even and 
isospin symmetric saddle point associated to (\ref{YX21}) hence (\ref{Y20}) 
belongs to the universality class described by the effective potential
($N_F=2$)
\be
{\cal L}_0 (T, \Phi )
= -h\Phi - \frac c{N_F} \Phi^{N_F} + \frac{g}{4}\Phi^4
\label{YX22}
\ee
with $\Phi \sim {\rm diag} \,\,\,P_{R,L}$. Eq.(\ref{YX22})
is only valid for temperatures $T\sim T_c\sim 1/4\Sigma$, with 
$c= 1-4\Sigma T$ and for a weak external field $h=-m$. The quadratic term  in
(\ref{YX22}), originates from the $2\times 2$ determinants in (\ref{YX21}) 
which are seen to break explicitly the $U_A(1)$ symmetry.

For three flavours we can just pursue the analogy with one flavour (\ref{X1})
and two flavours (\ref{YX21}) and write
\be
Z[T,\mu ] = &&\int \, dP\, dP^{\dagger}
e^{-N\beta\Sigma {\rm Tr}_f (PP^{\dagger})
+\theta N\beta ({\rm det}_f P + {\rm det}_f P^{\dagger} )}
\nonumber\\
&&\times \prod_{n=-\infty}^{+\infty} \,\,
{\rm det}_{2,f}^N \beta\left(  \matrix{i (m+P) & \omega_n + i\mu \cr
\omega_n + i\mu & i (m + P^{\dagger} ) \cr}\right)
\label{Y23}
\ee
where $P$ is a complex $3\times 3$ matrix. In the large $N$ limit, the 
$P$-integration can be again traded for a Grassmann integration, leading to
a three-flavour NJL model with a determinantal interaction. (\ref{Y23})
is $SU(3)_L\times SU(3)_R$ symmetric in the massless case. 
The QCD motivation for (\ref{Y23})
can be found in \cite{NJL,INS} and references therein.
In these models, we note that the additional parameter $\theta >0$ plays the 
role of the instanton density. 

If we were to expand the $U_A(1)$ breaking determinantal part of the 
exponential in (\ref{Y23}), rewrite each
determinant using new auxiliary Grassmann variables ($X_i^a$, $i=1,..,N_+$, and
$Y_j^b$, $j=1,..,N_-$, $a,b$ are flavor indices), and bosonize pairs
of opposite chirality as explained above (using ${\bf A}$'s instead of $P$'s)
then we can rewrite (\ref{Y23}) in the form
\be
Z[T, \mu ] = &&\sum_{N_{\pm}} \frac 1{N_+! N_- !} 
\left(\frac{{\theta N}\beta}{(i\beta)^3}\right)^{N_+ + N_-}\nonumber\\&&\times
\int \, d{\bf R} \,e^{-\frac{\beta\Sigma}2 \,{\rm Tr} ({\bf R}{\bf 
R}^{\dagger} )} \, {\rm det}\beta({\bf D} + {\bf R})
\label{YX23}
\ee
with the (rectangular) random matrix
\be
{\bf R} =
\left( \begin{array}{cccc}
        0 & {\bf A} & 0 & {\bf \Gamma}_R^{\dagger} \\
        {\bf A}^{\dagger} & 0 & {\bf \Gamma}_L^{\dagger} & 0\\
        0 & {\bf \Gamma}_L & 0 & {\bf \alpha} \\
        {\bf \Gamma}_R & 0 &{\bf \alpha}^{\dagger} & 0\end{array}\right)
\label{ST1}
\ee
and the sparse and deterministic (square) matrix
\be
{\bf D} = 
\left( \begin{array}{cccc}
        im & {\bf \Omega} +i\mu & 0 & 0 \\
        {\bf \Omega} +i\mu & im & 0 & 0\\
        0&0&0&0\\
        0&0&0&0\end{array}\right)
\label{ST2}
\ee
The block matrices above correspond to the following bosonized pairs
(the Matsubara indices connected to $q$'s are suppressed)
\be
{\bf A}^{ab}_{xy}
&=& \frac{1}{N\Sigma} q_{\!R x}^a {q^\dagger_{\!L\!}}^b_y \nonumber \\
{{\bf \Gamma}_{\!\!L}}^{\!ab}_{ix} = X_i^a {q_L^\dagger}^b_x \quad &,& \quad 
{{\bf \Gamma}_{\!\!R}}^{\!ab}_{jx} = {Y^\dagger_j}^a {q_R^b}_x \nonumber \\
\alpha^{ab}_{ij} &=&  X^a_i {Y^\dagger}_j^b
\label{ST3}
\ee
Each of the four rows and columns in the above matrices (\ref{ST1}-\ref{ST2})
correspond to a vector of length {\bf :}
${\bf 1}: 3\otimes N\otimes\infty$, ${\bf 2}: 3\otimes N\otimes\infty$, 
${\bf 3}: 3\otimes N_+$, ${\bf 4}: 3\otimes N_-$, where
$\infty$ stands for the number of Matsubara modes. We observe that 
(\ref{YX23}-\ref{ST3}) with the substitution $3\rightarrow 2$, hold
for two flavours as well, illustrating the model character of the present 
discussion.

The generic behaviour of these models occur at the critical points. For 
instance, (\ref{YX23}) undergoes a first order transition. 
At the critical point ($SU(3)_V$ symmetric phase) the universal 
potential for the parity even saddle point with $\Phi\sim {\rm diag} P$
follows from (\ref{Y1}),
with $N_F=3$, $h=-m$, $\mu^2= 4\Sigma T -1$, $c=4\theta T>0$ and $g=1/12 T^2$.
In the parameter range $(4\Sigma T -1 ) \geq 0$ and 
$(48\theta^2T^4-4\Sigma T + 
1 )\geq 0$, (\ref{Y1}) allows for a first order transition 
(weak external field) with a critical 
temperature $T_c$ fixed by the conditions ${\cal L}_0 ( T_c, \Phi ) ={\cal L}_0 
(T_c, 0 )$ and similarly for its derivative
${\cal L}'_0 ( T_c, \Phi ) ={\cal L}'_0 (T_c, 0 )$. As $T\rightarrow T_c$ from 
below, $\Phi  =  {2c}/{3g}$ with $\mu^2 = {2c^2}/{9g} \geq 0$. For strong
external fields, the transition turns second order.

The first order transition described by (\ref{Y1}) is reminiscent of 
the one encountered in the Maier-Saupe model for the transition from an 
isotropic molecular phase to a nematic phase 
in liquid crystals \cite{SAUPE,DEGENNES}. Indeed, 
if we were to denote by ${\cal Q}$ the order parameter (mean orientation of the 
molecules) in liquid crystals then in the mean-field approximation
the free energy per molecule is easily found to be \cite{STAT}
\be
{\cal G} =&& - T {\rm ln} 4\pi  + (1-0.4 T) {\cal Q}^2\nonumber\\
&&  -\frac 8{105 T^2} {\cal
Q}^3 + \frac {4}{175T^3} {\cal Q}^4 + ...
\label{YY24}
\ee
in analogy with (\ref{Y1}). The nematic phase corresponds to  molecules 
lined up along a preferential direction, but with no long-range order. This
phase is different from the smectic phase which is characterized by various 
degrees of translational ordering. We note that in the presence of an
external magnetic field (above quark masses) the first 
order transition becomes second order through a tricritical point. 

As a final point in this section, we note that in the mean-field approximation,
the equation of state $\Phi [ h]$ for $T\sim T_c$, allows for a simple 
understanding of the distribution of the quark eigenvalues $\nu (\lambda )$
near zero virtuality $\lambda \sim 0$  through
the identification 
$\pi \nu (\lambda ) = \, {\rm Re} \, \Phi (h = i\lambda + 0 )$, in the quenched
approximation. For two flavours 
the transition is second order with $\beta=1/2$, hence 
$\Phi [h] = h^{1/3}$, so that $\nu (\lambda ) \sim \lambda^{1/3}$
\footnote{From universality and Widom scaling, a similar argument would in
general suggest $\nu (\lambda )\sim \lambda^{\beta/(\beta+\gamma)}$ for
the distribution of eigenvalues, still in the quenched approximation. 
Such behaviour, however, calls for a non-local
equation of state as opposed to a local one in the Landau-Ginzburg analysis.}.
For three flavours the transition is first order for weak field
with $\nu (0) \sim 2c/3\pi g$ for $T<T_c$, and $\nu (0) \sim  0$ for 
$T>T_c$.

\vskip .5cm
{\bf 6. \,\,\,} The models we have discussed here  are schematic version 
of the NJL model and its relatives. For one flavour, the spectrum is 
that of constant $\sigma$ and $\pi$ (fluctuations in $P$) as well as 
constituent quarks with masses $\omega_*=(P_* + m)$. In confining theories, 
constituent quarks are barred from the low temperature part of the spectrum. 
Indeed, in two-dimensional QCD it can be shown that the quarks in light-cone
gauge satisfy the following dispersion relations \cite{THOOFT}
\be
E = -k + \frac {g_s^2}{\sqrt{2}\pi\lambda} {\rm sgn} (E - k)
\label{X15}
\ee
where $g_s^2$ is the fixed QCD coupling in large $N_c$ and 
$\lambda\rightarrow 0$ is an infrared regulator. For spatially constant modes, 
(\ref{X15}) reduces to $E= \pm g_s^2/\sqrt{2}\pi\lambda$. The spectrum 
collapses to two levels that are rejected to infinity when the infrared 
regulator is removed. Thermodynamics with the spectrum (\ref{X15}) yields
heavy Boltzmann penalty factors $e^{-1/T\lambda}$ or $e^{-\mu/\lambda}$. 
The contribution of the constant constituent quark modes drop  from the 
thermodynamics at low temperature or chemical potential, for
$\lambda\rightarrow 0$
\footnote{Zero modes are of course an exception. However, in matter they do not
sense the effect of temperature or density, except through a reorganization of 
the topological configurations.}. 
The mass gap  becomes infinite, and Fig. 1 would be
identically zero \cite{HANSSON}. As a result, the contribution to the pressure
at low temperature or chemical potential is hadronic. So the presently
discussed models do not reflect properly on confining theories.
They are, however, interesting to use around the critical point $T_c$, since 
the latter seems to be mainly driven by universality. In this regime, there may 
exist  a window of coexistence between just about to be freed massless 
quarks and long-wavelength mesonic correlations, depending on the nature and 
character of the transition.

\vskip .5cm
{\bf 7.\,\,\,} We have shown how standard chiral random matrix models can be 
understood in terms of two-level NJL models, whether in vacuum or matter. The
thermodynamics of these models is that of constituent quarks, with mean-field 
universality at zero chemical potential. Our analysis indicates that the 
large volume limit does not necessarily commute with the zero temperature limit 
in the presence of a finite chemical potential. This point may be relevant
for lattice simulations of QCD (quenched and unquenched)
at finite chemical potential \cite{MULATTICE}. In this respect, it would be 
interesting to reassess the Dirac spectrum at finite chemical potential in 
the presence of all Matsubara modes.

We have suggested non-standard chiral
random matrix models for two and three flavours,
with explicit $U_A(1)$ breaking as inspired by a number of effective
models to the QCD
vacuum. In large $N$ the character of the transition follows the general lore 
of universality \cite{PISARSKI}. In particular for three flavours, and for
a specific choice of the external parameters, the transition is first 
order and analogous to the isotropic-nematic transition encountered in 
liquid crystals.

\vglue 0.6cm
{\bf \noindent  Acknowledgments \hfil}
\vglue 0.4cm
This work was supported in part  by the US DOE grant DE-FG-88ER40388,
and by the Polish Government Project (KBN)  grant  2P03B19609. We thank
Gabor Papp for discussions. One of us
(IZ) thanks Andy Jackson, Misha Stephanov and Jac Verbaarschot for an 
informal discussion.

\vskip 1cm
\setlength{\baselineskip}{15pt}

\end{document}